%% file: main.tex
\documentclass[11pt,a4paper]{article}
\usepackage[utf8]{inputenc}
\usepackage[T1]{fontenc}
\usepackage{lmodern}
\usepackage[english]{babel}
\usepackage[margin=2.5cm]{geometry}
\usepackage{microtype}
\usepackage{graphicx}
\usepackage{xurl}
\usepackage[hidelinks]{hyperref}
\usepackage{bookmark}
\newenvironment{referencelist}{%
  \begin{list}{}{\setlength{\leftmargin}{1.5em}%
    \setlength{\itemindent}{-1.5em}\setlength{\itemsep}{0.45em}%
    \setlength{\parsep}{0pt}}%
}{\end{list}}
\title{Automated Disinformation and Malicious AI Swarms: Risks for Democracy and Development in Africa}
\input{authors}
\date{}
\hypersetup{pdftitle={Automated Disinformation and Malicious AI Swarms: Risks for Democracy and Development in Africa},pdfauthor={Daniel Thilo Schroeder et al.}}

\begin{document}
\maketitle
\clearpage

\begin{abstract}
\input{sections/00-abstract}
\end{abstract}

\input{sections/01-introduction-from-bots-to-swarms-in-the-global-south}
\input{sections/02-digital-ecosystems-and-democratic-fragility-in-africa}
\input{sections/03-citizen-journalism-and-the-weaponization-of-social-media}
\input{sections/04-from-botnets-to-malicious-ai-swarms}
\input{sections/05-case-study-1-mali-as-a-microcosm-of-emerging-threats}
\input{sections/06-case-study-2-ethiopia-as-a-laboratory-of-authoritarian-information-adaptation}
\input{sections/07-pathways-of-harms-to-democracy-in-africa}
\input{sections/08-governance-and-safeguards}
\input{sections/09-towards-resilient-information-ecosystems-for-democracy-and-development}
\input{sections/10-research-agenda-operationalising-ai-swarms-and-testing-their-effects}
\input{sections/11-conclusion}

\clearpage
\input{references}
\end{document}

%% file: authors.tex
\newcommand{\authormark}[2]{\mbox{#1\textsuperscript{#2}}}
\newcommand{\authoremail}[2]{#1: \href{mailto:#2}{\nolinkurl{#2}}}

\renewcommand{\maketitle}{%
  \begingroup
  \thispagestyle{empty}
  \centering
  \setlength{\parskip}{0pt}
  \vspace*{0.4cm}
  {\fontsize{19}{23}\selectfont\bfseries
    Automated Disinformation\par
    and Malicious AI Swarms:\par}
  \vspace{0.35em}
  {\fontsize{16}{20}\selectfont
    Risks for Democracy and Development in Africa\par}
  \vspace{1.1cm}
  {\fontsize{11}{17}\selectfont
    \authormark{Daniel Thilo Schroeder}{1}\quad
    \authormark{Philipp M. Lutscher}{2,3}\par
    \authormark{Samba Dialimpa Badji}{4}\quad
    \authormark{Stefan Brenner}{4}\quad
    \authormark{Wenchao Dong}{5}\par
    \authormark{Tsehaye Haidemariam}{6}\quad
    \authormark{Saheed Bidemi Ibrahim}{7}\quad
    \authormark{Amanuel Tesfaye Kebede}{8}\par
    \authormark{Jonas R. Kunst}{6,9}\quad
    \authormark{Johannes Langguth}{10}\quad
    \authormark{Pauline Lemaire}{11}\par
    \authormark{Mulatu Alemayehu Moges}{12}\quad
    \authormark{Lukasz Olejnik}{13}\par
    \authormark{Kristin Skare Orgeret}{4}\quad
    \authormark{Gerald Walulya}{14}\par}
  \vspace{0.8cm}
  {\fontsize{9.5}{12}\selectfont
    \renewcommand{\arraystretch}{1.05}
    \begin{tabular}{@{}r@{\hspace{0.5em}}l@{}}
      \textsuperscript{1} & SINTEF Digital, Oslo, Norway \\
      \textsuperscript{2} & Bavarian Research Institute for Digital Transformation (bidt), Munich, Germany \\
      \textsuperscript{3} & University of Bamberg, Bamberg, Germany \\
      \textsuperscript{4} & OsloMet--Oslo Metropolitan University, Oslo, Norway \\
      \textsuperscript{5} & Max Planck Institute for Security and Privacy, Bochum, Germany \\
      \textsuperscript{6} & BI Norwegian Business School, Oslo, Norway \\
      \textsuperscript{7} & Birmingham City University, Birmingham, United Kingdom \\
      \textsuperscript{8} & University of Helsinki, Helsinki, Finland \\
      \textsuperscript{9} & University of Oslo, Oslo, Norway \\
      \textsuperscript{10} & Simula Research Laboratory, Oslo, Norway \\
      \textsuperscript{11} & Chr. Michelsen Institute, Bergen, Norway \\
      \textsuperscript{12} & University of Agder, Kristiansand, Norway \\
      \textsuperscript{13} & Independent researcher / King's College London \\
      \textsuperscript{14} & Makerere University, Kampala, Uganda
    \end{tabular}\par}
  \vspace{0.9cm}
  {\fontsize{9}{11}\selectfont\textit{Author contact details}\par}
  \vspace{0.3cm}
  {\fontsize{8.5}{12}\selectfont\urlstyle{same}
    \renewcommand{\arraystretch}{1.15}
    \begin{tabular}{@{}l@{\hspace{1.4em}}l@{}}
      \authoremail{D. T. Schroeder}{daniel.t.schroeder@sintef.no} &
      \authoremail{P. M. Lutscher}{philipp.lutscher@bidt.digital} \\
      \authoremail{S. D. Badji}{sambadia@oslomet.no} &
      \authoremail{S. Brenner}{stefan.brenner@oslomet.no} \\
      \authoremail{W. Dong}{wenchao.dong@mpi-sp.org} &
      \authoremail{T. Haidemariam}{tsehaye.haidemariam@bi.no} \\
      \authoremail{S. B. Ibrahim}{saheed.ibrahim@mail.bcu.ac.uk} &
      \authoremail{A. T. Kebede}{amanuel.kebede@helsinki.fi} \\
      \authoremail{J. R. Kunst}{jonas.r.kunst@bi.no} &
      \authoremail{J. Langguth}{langguth@simula.no} \\
      \authoremail{P. Lemaire}{pauline.lemaire@cmi.no} &
      \authoremail{M. A. Moges}{mulatu_alem@yahoo.com} \\
      \authoremail{L. Olejnik}{me@lukaszolejnik.com} &
      \authoremail{K. S. Orgeret}{kristin.orgeret@oslomet.no} \\
      \multicolumn{2}{c}{\authoremail{G. Walulya}{gerald.walulya@mak.ac.ug}}
    \end{tabular}\par}
  \endgroup
}

%% file: sections/00-abstract.tex
Generative artificial intelligence is reshaping how information is produced, accessed, and circulated, while also enabling disinformation campaigns of increasing scale and sophistication. Although there is currently no clear evidence that fully autonomous ``AI swarms'' are conducting influence operations at scale, their enabling components, including large language models, multi-agent coordination, and real-time optimization, are advancing rapidly and becoming more accessible. Building on existing functional accounts, we define malicious AI swarms as coordinated, persistent, and adaptive multi-agent systems designed to conduct influence operations and distinguish them from AI-assisted content production and centrally managed networks of synthetic personas. We then examine the implications of these developments for hybrid regimes and conflict-affected states in Africa, where limited institutional and regulatory capacity and fragile media ecologies may heighten vulnerability. Specifically, we consider how automated influence operations could infiltrate online communities, fabricate consensus, and erode trust in governance and development processes. Drawing on two case studies of Mali and Ethiopia, we examine how these risks interact with conflict, contested legitimacy, polarized digital publics, diaspora participation, and different configurations of state and non-state influence. Mali illustrates a fragmented information ecology shaped by competing actors and weak gatekeeping, while Ethiopia shows how digital influence can also become integrated into more organized state-led strategies of narrative management and political mobilization. We also examine how African-language and training-data asymmetries, together with hybrid human--AI operations, may affect both influence capabilities and defensive responses. The analysis develops a near-future risk scenario in which AI-driven coordination becomes accessible to state and non-state actors in conflict-affected settings. Finally, we propose a layered governance approach spanning data, models, agents, platforms, civic institutions, and regional coordination, and outline safeguards for democratic participation, peacebuilding, and development across diverse African contexts.

%% file: sections/01-introduction-from-bots-to-swarms-in-the-global-south.tex
\section{Introduction: From Bots to Swarms in the Global South}
\label{sec:introduction-from-bots-to-swarms-in-the-global-south}

Rapid advances in generative artificial intelligence are reshaping the global information environment and transforming how influence is produced, distributed, and perceived. Following the information-disorder framework, we distinguish disinformation from misinformation by emphasizing intent, while recognizing that both can dynamically circulate through the same social and technical channels (Wardle \& Derakhshan, 2017). For foreign or cross-border operations, we also draw on the EU concept of foreign information manipulation and interference (FIMI), which focuses on intentional, coordinated patterns of manipulative behavior rather than solely on the truthfulness of individual content (European External Action Service {[}EEAS{]}, 2023). Because this paper also examines domestic and hybrid actors, we treat FIMI as a subset of influence operations rather than as its overarching frame. While fully autonomous swarm networks remain an emerging threat (Orlando et al., 2026), the technical building blocks -- generative text, agentic planning, and multi-modal synthesis -- are already available (Wang et al., 2024). Large language models and multi-agent simulations demonstrate components that could enable automated campaigns to coordinate synthetic personas and adapt to feedback (Orlando et al., 2026; Olejnik et al., 2026). As Schroeder et al. (2026) argue, these malicious AI swarms represent a qualitative leap beyond traditional botnets, as they possess autonomy, memory, and the ability to simulate coherent communities rather than merely amplifying slogans. Building on this functional account, we develop an operational and empirically testable definition of malicious AI swarms, distinguishing their constitutive properties---multiplicity, persistent roles or identities, coordination, adaptation, and collective task allocation---from the capabilities and harms they may produce. We further distinguish AI-assisted content production and coordinated synthetic personas from adaptive multi-agent campaigns, while treating goal-revising swarms as a speculative upper end of the capability spectrum.

The potential maturation of these technologies signals a new phase of information warfare in which realistic synthetic audio and video can enable more convincing deception (Chesney \& Citron, 2019). While these dynamics pose a challenge to established democracies, their risks are profoundly asymmetric. In many Global South contexts, automated disinformation intersects with broader development constraints, including institutional fragility, uneven digital infrastructure, and limited regulatory capacity, potentially heightening exposure to destabilizing influence operations (Okolo, 2025). Across African countries, expanding mobile connectivity and social-media use have reshaped how many citizens obtain and exchange critical information, although adoption remains highly uneven. The International Telecommunication Union (ITU) estimated that 36\% of the population in its Africa region used the Internet in 2025, compared with 74\% globally (International Telecommunication Union {[}ITU{]}, 2025). Radio remained the most widely used news source across the 38 countries surveyed by Afrobarometer in 2024--2025 (Conroy-Krutz et al., 2026). This partial connectivity limits the immediate direct reach of automated influence operations but also creates a window for establishing safeguards before connectivity and AI-enabled influence capabilities expand further.\footnote{That said, (false) online information and rumours often spread through communities, i.e. the well connected act as gatekeepers} Among connected publics, social-media, and messaging platforms have become important spaces for news, political discussion, and campaign organization. WhatsApp has become an important channel for political organization and mobilization in several African contexts. Research documents its use in party election campaigns in Ghana and Nigeria (Fisher et al., 2024) and in pro-opposition political organizing in Zimbabwe (Young, 2023). These examples should not be read as evidence of a single African information environment: the political effects of digital platforms vary with regime type, conflict dynamics, platform use, linguistic diversity, institutional trust, diaspora participation, and the relationship between online and offline communication.

Mali and Ethiopia provide two analytically related but institutionally distinct contexts in which digital communication has become deeply entangled with political conflict, contested legitimacy, and struggles over information control. In both cases, social-media infrastructures have developed under conditions of armed conflict, restricted media environments, polarized political discourse, and low or uneven trust in institutions. Online communication has also become part of broader forms of participatory propaganda and political mobilization in which domestic actors, diaspora communities, state institutions, and other organized groups contribute to struggles over legitimacy and territorial authority. These shared conditions make Mali and Ethiopia particularly interesting for examining how emerging AI-enabled influence operations could interact with already contested information environments. At the same time, the cases differ in important ways: Mali illustrates a more fragmented conflict ecology shaped by weak gatekeeping and competing domestic and foreign actors, whereas Ethiopia demonstrates stronger state capacity to organize and mobilize digital influence in support of regime narratives and foreign-policy objectives itself. In both countries, these dynamics are further shaped by linguistic diversity, uneven connectivity, diaspora participation, and the interaction between online communication and offline political mobilization: factors that are also frequently encountered in other countries in the Global South.

The rise of citizen journalism across Africa has further complicated this media landscape. In contexts where professional journalism is underfunded or restricted, citizens have increasingly assumed roles once reserved for trained reporters -- documenting events, verifying claims, and shaping public narratives. As de Bruijn et al. (2025) show in their study of the Malian Twittersphere, these grassroots information producers operate within what they call networked journalism: a hybrid ecosystem where bloggers, activists, government officials, and NGOs collaboratively construct and contest meaning (see also Bosch, 2017; Mutsvairo \& Salgado, 2022). Yet this participatory dynamism is double-edged. The same infrastructures that empower citizens to report from conflict zones can also be weaponized to amplify disinformation, hate speech, online harassment, propaganda, or FIMI. In Mali, networked journalism often blurs into networked conflict, as online exchanges mirror---and sometimes escalate---offline tensions. Ethiopia similarly demonstrates how digital participation can become entangled with conflict mobilization, strategic narratives, diaspora engagement, and organized political campaigning. In both settings, the boundary between civic participation and coordinated influence becomes increasingly porous, although the organizational forms and degree of state involvement differ substantially.

Advances in generative AI, building on foundational developments in transformer architectures (Vaswani et al., 2017), together with more recent progress in autonomous agent systems (Wang et al., 2024), are transforming these dynamics at increasing speed and scale. What were once troll-farm influence operations and automated political accounts (e.g., Eady et al., 2023; Bolsover \& Howard, 2019) may increasingly evolve into coordinated networks of AI agents capable of maintaining persistent identities, learning from interaction, and tailoring messages to different audiences. As described by Schroeder et al. (2026), such malicious AI swarms could infiltrate online communities, sustain coherent narratives across thousands of synthetic accounts, and adapt their discourse to maximize engagement with minimal human supervision. Recent experimental work indicates that small, open-weight language models running on commodity hardware can reduce the time and cost of generating short persona-conditioned political content (Olejnik, 2026; Olejnik et al., 2026). Their use in coordinated real-world campaigns, however, remains untested. Therefore, we anticipate that hybrid human--AI influence operations will precede fully autonomous swarms, making them the most immediate concern. Human strategists, account operators, influencers, and local-language specialists can supply contextual knowledge, credibility, platform access, and strategic judgement, while generative systems provide scale, translation, message variation, and support for persistent personas (Kunst, Bierwiaczonek, et al., 2026). In one documented state-backed campaign, AI adoption coincided with greater article output without a detected loss of credibility in a survey experiment (Wack et al., 2025). A key unit of analysis is thus the wider sociotechnical network through which human and computational actors jointly organize influence, rather than the autonomous agent in isolation.

Across Africa social media often substitutes for public deliberation, and the consequences could be profound. AI-driven campaigns can exploit linguistic diversity, fragile trust in institutions, and existing social cleavages to fabricate consensus, inflame polarization, or erode confidence in elections and governance. In this sense, automation could magnify preexisting vulnerabilities, turning local information fragility into wider democratic risk (Okolo, 2025). These risks also threaten policy implementation and, by extension, development outcomes: when information systems lose credibility, citizens may become less willing to follow public-health guidance, support security-sector reforms, trust service-delivery commitments, or participate in peacebuilding processes. Even well-designed policies can lose legitimacy before they take effect if the public sphere in which they are explained and contested has already been distorted by coordinated manipulation. The risks and responses examined here therefore bear directly on the Sustainable Development Goals, most clearly SDG 16's commitments to accountable institutions, inclusive decision-making, and public access to information, but also SDG 9's focus on equitable digital infrastructure and SDG 4's emphasis on education for sustainable development (United Nations, 2015).

A systematic review identifies a strong focus on U.S.-centric events and limited attention to African contexts in English-language research on AI and disinformation (Grub \& Humprecht, 2025). Generative and agentic AI raises further questions about these systems' interactions with African political and media environments (Okolo, 2025). In this paper, we address this gap by integrating insights from computational social science, media studies, political science, ethnographic research and AI governance. Accordingly, we offer a forward-looking risk analysis, anticipating how malicious AI swarms could operate in African political contexts before they materialize at scale. Our paper makes four related contributions. First, it develops an operational definition of malicious AI swarms and identifies graded dimensions through which they can be studied empirically. Second, it analyses how African-language and training-data asymmetries may shape both influence capabilities and defensive limitations. Third, it conceptualizes hybrid human--AI influence operations as distributed sociotechnical systems. Fourth, it develops a layered governance framework spanning data, models, agents, platforms, civic institutions, and regional coordination.

To ground this analysis, we use Mali and Ethiopia as two illustrative cases in the Global South. The comparison of Mali and Ethiopia lies in combining several shared conditions---armed conflict, contested legitimacy, polarized digital publics, diaspora participation, and constrained media environments---while differing in the organization of influence and state capacity. This contrast allows us to examine both common mechanisms and context-specific pathways through which AI-enabled influence operations may develop. We aim not only to describe a potential threat but also to clarify its implications for development and governance, including how automated disinformation could distort public deliberation, undermine peacebuilding, and weaken the institutional foundations of democratic development (McCoy et al., 2018; Okolo, 2025).

The remainder of this article proceeds as follows. Section~\ref{sec:digital-ecosystems-and-democratic-fragility-in-africa} examines Africa's rapidly evolving digital ecosystems, tracing how connectivity, platform dependence, and institutional fragility create conditions conducive to automated disinformation. Section~\ref{sec:citizen-journalism-and-the-weaponization-of-social-media} builds on insights from networked journalism, networked conflict, and participatory propaganda to analyse how conflict-affected information environments blur the boundary between civic expression, political mobilization, and coordinated manipulation. Section~\ref{sec:from-botnets-to-malicious-ai-swarms} introduces malicious AI swarms, providing a technical definition, capability taxonomy, and analysis of hybrid human--AI operations and linguistic asymmetries. Sections~\ref{sec:case-study-1-mali-as-a-microcosm-of-emerging-threats} and~\ref{sec:case-study-2-ethiopia-as-a-laboratory-of-authoritarian-information-adaptation} develop the Mali and Ethiopia case studies. Section~\ref{sec:pathways-of-harms-to-democracy-in-africa} synthesizes these insights into pathways of harm to democracy and development, while Section~\ref{sec:governance-and-safeguards} develops a layered governance approach. Section~\ref{sec:towards-resilient-information-ecosystems-for-democracy-and-development} considers the institutional and civic conditions for resilience. Finally, Section~\ref{sec:research-agenda-operationalising-ai-swarms-and-testing-their-effects} sets out a research agenda for operationalizing swarm-like behaviour and testing its effects, before the conclusion in Section~\ref{sec:conclusion}.

%% file: sections/02-digital-ecosystems-and-democratic-fragility-in-africa.tex
\section{Digital Ecosystems and Democratic Fragility in Africa}
\label{sec:digital-ecosystems-and-democratic-fragility-in-africa}

The expansion of digital communication across Africa has unfolded under structural conditions that differ considerably from those in more established media environments. Mobile internet use has expanded in Sub-Saharan Africa alongside improvements in network infrastructure, although affordability remains a major barrier (GSMA, 2024; ITU, 2023). For connected citizens, mobile internet and social-media platforms have become increasingly important gateways to news, political discourse, and public debate. Yet access remains uneven: mobile internet use is far less prevalent among rural and lower-income communities, and radio, television, community media, interpersonal networks, and word-of-mouth continue to mediate how online information circulates offline (Kokutse \& Thompson, 2024; de Bruijn et al., 2025; Nyabola, 2018). Social media platforms such as WhatsApp, Facebook, X/Twitter, and TikTok thus serve not merely as supplementary information channels for connected publics, but also as central arenas for civic communication and collective sense-making. This rapid digital uptake, however, has not been accompanied by comparable growth in institutional oversight, media literacy, or infrastructure for information verification.

The uneven geography of connectivity shapes how far, and to whom, such fabricated consensus travels. Because internet access correlates with income, education, and urban residence (Shanahan \& Bahia, 2024; ITU, 2023), swarm-generated narratives would initially reach connected audiences more directly rather than the population at large. Far from limiting the harm, this skew may concentrate it. The connected minority overlaps substantially with those who set the agenda of national debate: journalists, civil servants, politicians, teachers, religious leaders, diaspora commentators, and the urban middle class whose posts feed radio talk shows and word-of-mouth networks (Brinkerhoff, 2009; de Bruijn et al., 2025; Nyabola, 2018). The two-step flow of communication (Katz \& Lazarsfeld, 1955) offers a framework for considering how manipulated online discourse may be relayed through offline intermediaries. In such transfers, contextual cues that might reveal synthetic origins could be lost. The result is a form of epistemic inequality. Better-connected citizens may disproportionately shape the online narrative record, including its manipulated portions, while people with more limited access may encounter these narratives indirectly through interpersonal and offline media channels (Gondwe et al., 2026). This asymmetry carries at least three consequences. First, observers who treat social media as a barometer of public sentiment, including journalists, election monitors, and policymakers, risk mistaking synthetic consensus among the connected for national opinion. Second, where connectivity maps onto region, ethnicity, or partisan alignment, targeted campaigns can mobilize or demobilize specific constituencies while remaining largely invisible to others. Third, corrections delivered through online fact-checks or platform labels may not follow falsehoods across the digital divide. At the same time, partial reach also bounds the threat: as long as large parts of the population remain offline, swarm influence is refracted through offline institutions and interpersonal trust, underscoring the mediating role of community media and local intermediaries discussed below.

African media ecosystems have long been shaped by constraints on press freedom, under-resourced news organizations, and uneven state capacity (Wasserman \& Madrid-Morales, 2022). In such contexts, social media often functions as a substitute for independent journalism, providing citizens with sources of reporting that bypass official narratives. As shown in research on the Malian Twittersphere (de Bruijn et al., 2025), social networks in conflict-affected countries evolve in unstable and contested informational environments, where journalists, bloggers, activists, state actors, and diaspora communities all contribute to shaping public discourse. Digital participation, therefore, emerges within media ecologies already marked by political tension and distrust---conditions that make informational cohesion difficult to sustain.

Diaspora networks further complicate these dynamics. Digital diasporas enable transnational political engagement (Brinkerhoff, 2009), and diaspora accounts contributed substantial posting volume in the Malian Twittersphere studied by de Bruijn et al. (2025). While this transnational engagement can strengthen pluralism, it can also amplify polarized narratives or introduce external political agendas. In the Malian case, for example, de Bruijn et al. (2025) describe how citizen journalists and social-media activists operate in a state of networked conflict, where online narratives reflect and may contribute to tensions on the ground. In Ethiopia, analysis of Facebook discourse during conflict identifies hostile rhetoric that may reinforce ethnopolitical polarization (Gessesse \& Moges, 2026). Digital publics thus become entangled with political alignments, identity contestations, and evolving conflict dynamics.

These structural conditions can collectively heighten susceptibility to coordinated manipulation. Studies from Ethiopia and Mali show that pressure on fact-checkers and difficulty accessing reliable sources can redirect verification work away from sensitive political claims (Badji et al., 2024), while journalists report competing understandings of disinformation and a lack of established truth-telling institutions (Mutsvairo et al., 2026). Within the platform-dependent and transnational media environments described above, such weaknesses can make coordinated manipulation harder to identify and counter (Africa Center for Strategic Studies, 2024; Wasserman \& Madrid-Morales, 2022). The health of democratic deliberation in many African states is therefore increasingly tied to the integrity of their digital communication ecosystems.

This fragility matters profoundly in a moment of accelerating technological change through the rapid deployment of AI (Okolo, 2025). Understanding the socio-political risks posed by increasingly autonomous disinformation systems, therefore, requires situating them within these specific infrastructural, political, and communicative conditions.

%% file: sections/03-citizen-journalism-and-the-weaponization-of-social-media.tex
\section{Citizen Journalism and the Weaponization of Social Media}
\label{sec:citizen-journalism-and-the-weaponization-of-social-media}

The expansion of social media in Africa has enabled new forms of civic participation, where individuals increasingly assume functions once associated with professional journalism (Mutsvairo \& Salgado, 2022). In environments marked by limited press freedom, weak media institutions, and geographic barriers to traditional reporting, citizens have turned to digital platforms to document events, verify claims, and circulate political narratives. This bottom-up production of information has been widely interpreted as a democratizing force, offering alternative channels of accountability and enabling marginalized groups to participate in public debate (Bosch, 2017).

However, in the Malian context, this participatory media environment has evolved into networked journalism. Rather than a clear distinction between ``journalists'' and ``audiences,'' the Malian Twittersphere brings together diverse content producers who interpret and reframe political conflict, while also displaying concentrations of state-aligned influence (de Bruijn et al., 2025). Authority is negotiated through institutional position as well as social connections, follower networks, and perceived authenticity.

This decentralization of narrative production is double-edged. On the one hand, it expands access to public discourse and enables local perspectives to surface in ways that state media or international outlets often overlook. On the other hand, it dilutes verification norms and weakens mechanisms of editorial accountability. In settings where trust in government, media, and international actors is already strained, claims can circulate without institutional validation and rapidly take hold as shared belief. Emotional resonance, identity alignment, and viral affordances often outweigh evidentiary credibility in determining what becomes socially accepted as truth (Nyabola, 2018).

These dynamics are particularly salient in contexts of conflict and political instability. As shown in Mali and comparable conflict-affected information environments, digital platforms become arenas where political actors and citizen reporters compete to define the meaning of unfolding events (de Bruijn et al., 2025; Africa Center for Strategic Studies, 2024). Digital communication becomes a domain of networked conflict, where online debates influence attitudes, mobilization, and collective memory. Rumor, accusation, and calls to action may circulate before claims can be independently verified. Research on hashtag dynamics during Ethiopia's Tigray conflict shows how hashtags can become ``hatetags'': emotionally charged vehicles for hate speech, dehumanizing rhetoric, and conflict escalation, reinforcing the need to treat digital platforms as active conflict infrastructures rather than passive channels (Orgeret et al., 2026).

Such conditions create a fertile environment for strategic manipulation. Political elites, armed groups, foreign actors, and commercial disinformation networks have learned to exploit citizen journalism ecosystems by seeding narratives through trusted local voices, amplifying divisive commentary, or mimicking the stylistic cues of grassroots reporting (Africa Center for Strategic Studies, 2024). The absence of clear boundaries between genuine civic participation and orchestrated influence makes detection difficult and contestation politically charged. As a result, the very infrastructures that enable democratic expression can be weaponized to intensify polarization and undermine trust.

In this context, African regimes have also been important actors manipulating and weaponizing digital media for their own purposes (Mare, 2020). This includes shutting down internet access during elections or periods of conflict and instability (Freyburg \& Garbe, 2018; Chala \& Workneh, 2026), conducting campaigns using social media (Bosch et al., 2020), and organizing digital media armies to influence online discourses (Pohjonen et al., 2026b). States have demonstrated they can adapt and adopt new tactics quickly: government-related social media accounts engaged in coordinated inauthentic behaviour in Uganda during the 2021 general elections, and when these accounts were suspended by Facebook, access to the social media service was subsequently blocked (Lemaire \& Karyeija, 2024). States adjust internet shutdowns to ongoing events, using measures ranging from internet blackouts and network shutdowns to platform blocks and throttling (Chala \& Workneh, 2026). State-linked accounts also use social media to propagate their narratives (Kebede \& Pohjonen, 2025). Comparative research documents state-associated uses of generative AI (Beacken et al., 2026), while a report on Rwanda identifies an AI-empowered pro-government influence network on X (Wack et al., 2024).

This ambiguous, distributed, and emotionally charged communication landscape, filled with actors willing and capable of instrumentalizing it to achieve their political objectives, is the informational terrain into which more advanced forms of automated disinformation -- such as AI-assisted coordinated campaigns -- would intervene (Goldstein et al., 2023; Wang et al., 2024). The vulnerabilities are not isolated technical flaws but outcomes of the social structure of digital public life: low verification thresholds, blurred authority, identity-based trust, and platform incentives that reward contention over deliberation (Nyabola, 2018). These systemic features heighten the stakes of introducing increasingly autonomous and adaptive disinformation systems.

%% file: sections/04-from-botnets-to-malicious-ai-swarms.tex
\section{From Botnets to Malicious AI Swarms}
\label{sec:from-botnets-to-malicious-ai-swarms}

Coordinated disinformation campaigns have a documented history across several African political contexts (Alberts, 2025; Nanfuka et al., 2024). Coordinated accounts have long formed part of broader influence infrastructures in several African information environments, operating alongside coordinated fake accounts, paid influencers, political consultancies, pseudo-news outlets, and locally recruited or co-opted operators (Africa Center for Strategic Studies, 2024; Graphika \& Stanford Internet Observatory, 2020). Such hybrid networks have been documented in domestic election campaigns (Meta, 2021b) and foreign information operations (Huntley, 2022; Meta, 2021a). Human labour and coordination remain important operational requirements, which generative models may partially reduce (Goldstein et al., 2023). Detection systems on major platforms can seek indicators of such campaigns by their uniform posting rhythms, duplicated content structures, or shared metadata traces (Goldstein et al., 2023).

Recent advancements in generative artificial intelligence, however, are altering these constraints. Large language models can generate human-like text across multiple languages and dialects, respond dynamically to conversation contexts, and imitate emotional tone and local idioms. When embedded in multi-agent frameworks, these models could support persistent identities, track interaction histories, and vary message timing, voice, and framing (Wang et al., 2024). The underlying components of malicious AI swarms (Schroeder et al., 2026) -- memory, adaptive planning, persona consistency, and decentralized communication -- are therefore already emerging, even if fully autonomous, self-reinforcing swarm networks are not yet proven to be widely deployed. The key claim is not that these swarms are already a routine feature of African elections or conflicts, but that their component capabilities are converging in ways that lower the threshold for future experimentation by state and non-state actors.

To apply malicious AI swarms comparatively and empirically across African information ecosystems, however, it is necessary to distinguish operational capabilities from the constitutive properties that make an AI-enabled influence system a swarm. Following Schroeder et al. (2026)'s work, we define a malicious AI swarm as a coordinated system of multiple AI-enabled agents that maintain partially persistent identities or operational roles, act on shared or mutually relevant information, adapt their behaviour in response to environmental feedback, and collectively pursue an influence objective with some degree of operational autonomy. Note that the defining feature is not scale or the use of generative content, but the combination of multiplicity, coordination, persistence, adaptation, and collective agency. We therefore distinguish malicious AI swarms from related phenomena such as automated botnets, coordinated synthetic personas, and human-led influence campaigns that use generative AI.

Five dimensions are especially important for distinguishing a swarm from previous-era forms of automation. Multiplicity refers to the presence of multiple agents or synthetic personas rather than a single content-generation system. Coordination means that the agents' actions are organized around a shared campaign objective, even where individual messages or personas appear independent. Persistence refers to the maintenance of identity, memory, interaction history, or assigned roles across time. Adaptation requires that agents modify their behaviour in response to user engagement, platform conditions, counter-messaging, or changing political events. Collective agency refers to the capacity of the system to distribute tasks, share relevant information, revise tactics, or reallocate effort with less continuous human direction than conventional influence operations require, reflecting a broader shift from externally specified coordination toward increasingly goal-directed and adaptive forms of computational agency (Haidemariam, 2026). These dimensions may be present to different degrees and should not be treated as binary properties.

This definition also clarifies what does not, by itself, constitute a malicious AI swarm. A large volume of AI-generated political content is not necessarily a swarm. Nor is a set of synthetic accounts that merely reposts centrally produced messages according to a fixed schedule. Such systems may be automated and coordinated, but they lack the adaptive, persistent, and collectively organized characteristics associated with swarm-like behaviour. Similarly, human operators using generative AI to manage multiple accounts may constitute an AI-assisted influence operation without reaching the threshold of an autonomous or semi-autonomous swarm.

To avoid treating all forms of AI-enabled influence as equivalent, these systems can be organized along a four-level capability ladder. The first level is AI-assisted content production, in which human operators use generative models to create, translate, summarize, or vary political messages. The second level is coordinated synthetic personas, in which multiple accounts maintain differentiated identities and distribute strategically aligned content, but remain substantially directed by human operators. The third level is adaptive multi-agent campaigning, in which agents respond to engagement signals, alter messages or targets, and coordinate actions across accounts with reduced human supervision. The fourth level is goal-revising malicious swarms, in which agents not only adapt tactics but also modify intermediate objectives, redistribute tasks, and sustain a campaign through recursive feedback loops.

This distinction is important because coordination should not be conflated with autonomy, and message variation should not be equated with goal-directed reasoning. Haidemariam (2026) conceptualizes the transition from externally specified optimization to goal-directed agentic reasoning as an ``agentic turn'' in which systems increasingly maintain, evaluate, and revise goals through recursive processes. Applied to influence operations, this suggests that the threshold for a genuine swarm lies not merely in producing many messages, but in coordinating multiple agents that can respond to changing conditions and collectively sustain an influence strategy.

Claims about multilingual adaptability also require qualification. The ability of a language model to produce grammatically plausible text is not equivalent to the ability to communicate persuasively within a specific political, historical, and cultural context. Recent evaluations show substantial and uneven performance gaps between English and other African languages across classification, reasoning, knowledge, and generative tasks. These gaps reflect longstanding disparities in NLP resource availability and research attention (Joshi et al., 2020; Ojo et al., 2025). A broader review similarly identifies limited language and script coverage, tokenization problems, data scarcity, and inadequate evaluation as continuing barriers to reliable African-language model performance (Hussen et al., 2026).

These limitations create a double asymmetry. On the one hand, weak linguistic and contextual performance may constrain fully autonomous influence campaigns in underrepresented languages. On the other hand, the same scarcity of data and evaluation resources can make automated moderation, fact-checking, attribution, and coordinated-behaviour detection less effective in those languages. Linguistic underrepresentation may therefore reduce some offensive capabilities while simultaneously weakening defensive capacity. Attackers could also circumvent model limitations through code-switching, translation into dominant regional or colonial languages, multimodal content, or collaboration with local human intermediaries. Consequently, low-resource languages should not be treated simply as protective barriers.

Training-data composition is central to this asymmetry. Analyses of training-data samples identify linguistic imbalance, limited source transparency, and uneven representation of genres and periods (Haidemariam \& Gran, 2026). Political information environments also matter: governments exercising tighter control over domestic media are structurally better positioned to shape the material entering training corpora. Waight et al. (2026) identify Chinese state-coordinated media in training corpora and show that additional pretraining on it can make open-weight models more favorable toward Chinese institutions and leaders. Separate audits identify language differences and a cross-national association with media control. Beyond this structural pathway, state and state-aligned actors have also attempted to flood web sources with strategically framed content intended to shape search and chatbot outputs. This practice is often described as LLM grooming and may constitute training-data poisoning when manipulated content enters model-training pipelines (Dylan \& Grossfeld, 2025; Wilson, 2026). For influence operations, these deficiencies can produce factual and cultural errors that make synthetic personas easier to identify. However, where the same deficiencies affect defensive systems, they can also generate false positives, overlook locally meaningful hate expressions, or fail to recognize coordinated narratives expressed through vernacular, coded, or culturally specific language. The relevant question is therefore not merely whether a model ``supports'' an African language, but which communities, registers, scripts, and political contexts are represented in its training and evaluation data.

Although fully autonomous malicious AI swarms remain primarily experimental rather than widely documented in operational influence campaigns, the technical and economic barriers to their deployment are falling rapidly. Recent experimental work on AI propaganda factories shows that small, locally run language models can generate short persona-conditioned political messages and automate evaluation on commodity hardware, reinforcing that influence-content manufacturing is becoming an operational risk (Olejnik, 2026; Olejnik et al., 2026). A more immediate risk may be hybrid or ``cyborg'' propaganda rather than fully autonomous swarms. In such operations, human participants and generative systems perform complementary functions. AI systems can rapidly generate message variants, summarize events, translate material, maintain draft personas, and test alternative emotional or ideological frames. Human operators can provide strategic judgement, local language competence, account access, cultural knowledge, and the credibility associated with an established social identity. Verified accounts, political influencers, journalists, activists, diaspora commentators, or apparently ordinary citizens can then select, modify, ratify, and redistribute AI-generated material, allowing a centrally supported campaign to appear as decentralized grassroots expression (Kunst, Bierwiaczonek, et al., 2026).

Synthetic personas can be constructed through combinations of identity cues rather than through language alone. Names, profile images, clothing, accent, biographical details, regional references, religious symbols, interaction styles, and expressed political memories can be systematically assembled to signal membership in a particular community. A framework for synthetic identity treats AI-generated identities as multidimensional bundles of cues that may shape categorization, perceived similarity, threat, empathy, and social distance (Haidemariam \& Zheng, 2026). Transposed to social-media influence operations, this suggests that malicious actors could design personas to appear as culturally and politically recognizable in-group members. Such perceived in-group membership may be particularly consequential in homophilic clusters, where interactions and information diffusion are biased toward like-minded users (Cinelli et al., 2021). Detection systems should therefore examine the provenance and consistency of multimodal identity cues, rather than focusing only on textual similarity or posting frequency.

Experiments demonstrate low-cost production of political content using locally run models, although this does not establish the cost of a complete influence campaign (Olejnik, 2026; Olejnik et al., 2026). In conflict-affected African states, the most immediate risk may come from domestic governments and their challengers using these tools to shape conflict narratives. External actors may also conduct campaigns or support local partners with resources and expertise (Graphika \& Stanford Internet Observatory, 2020; Huntley, 2022; Meta, 2021a). Effective operations still depend on locally embedded networks, making hybrid human--AI campaigns more likely than fully autonomous ones. Simulations combine scale with heterogeneous agent behaviour (Olejnik et al., 2026), suggesting the possibility of campaigns that appear organic while remaining coordinated.

In a plausible scenario, malicious AI swarms could infiltrate and shape discourse inside politically salient online communities. Unlike traditional botnets, which simply broadcast slogans or retweet propaganda, swarms could engage in deep conversations with users, respond to criticism, adjust narrative strategies, and cultivate credibility over time (Schroeder et al., 2026). They could map the structure of social networks, identify influential accounts, and tailor messages to resonate with specific ethnic, religious, or regional identities. In divided societies, where trust is relational rather than institutional, the ability to mimic the voice of ``someone like me'' could amplify influence. This is a proposed AI-enabled extension of the social dynamics discussed by Nyabola (2018).

These developments matter because they intersect with the structural vulnerabilities outlined in the previous sections. In environments where verification norms are weak, editorial authority is decentralized, and public trust is fragile, the introduction of adaptive, conversational, and identity-aware disinformation systems poses risks far greater than those associated with earlier automation. The danger does not lie only in the volume of false content but in the ability to engineer the appearance of consensus -- to make certain viewpoints seem widely held, locally grounded, and socially validated even when they are algorithmically manufactured (Schroeder et al., 2026).

Understanding these emerging capabilities is therefore essential for assessing how automated disinformation may shape the trajectories of democratic participation, conflicting dynamics, and developmental governance across African contexts.

%% file: sections/05-case-study-1-mali-as-a-microcosm-of-emerging-threats.tex
\section{Case Study 1: Mali as a Microcosm of Emerging Threats}
\label{sec:case-study-1-mali-as-a-microcosm-of-emerging-threats}

Mali offers a particularly revealing context for examining how digital ecosystems, citizen journalism, and emerging AI-based disinformation capabilities intersect. The country has experienced recurring political crises and an ongoing security conflict involving state forces, armed groups, and foreign actors (Freedom House, 2024; Human Rights Watch, 2024). In this environment, social media has become a central arena for interpreting events, asserting legitimacy, and mobilizing public sentiment. Because traditional media infrastructures are limited and journalists often operate under threat, online platforms provide important spaces for connected publics to discuss political events (de Bruijn et al., 2025; Bruls \& de Bruijn, 2025). Network shutdowns have also been documented in Mali (Rydzak et al., 2020).

The structure of the Malian Twittersphere reflects these tensions. Schroeder et al. (2023) document a follower-network snapshot containing 56,505 accounts identified as Malian, including bloggers and political actors. Within this wider media environment, social media blur conventional boundaries between reporting, commentary, and political advocacy (de Bruijn et al., 2025). Survey evidence also indicates substantial pressure on journalistic gatekeeping: approximately 68\% of the journalists and media professionals in Dembele et al.'s (2024, p. 281) quota-sampled survey reported concerns about censorship or practiced self-censorship, particularly around security and justice issues. Under these conditions, claims may gain authority through circulation, repetition, and alignment with group identities before they can be independently verified.

Citizen journalism plays a prominent role in this ecosystem. Individuals and informal media networks report on conflict events, territorial shifts, military casualties, and diplomatic developments (de Bruijn et al., 2025). While this creates pathways for transparency and local voice, it also introduces significant risks. Real-time reporting from conflict zones commonly relies on unconfirmed accounts, unverifiable images, and interpretive narratives shaped by political alignment. These conditions could increase the risks posed by AI-assisted disinformation. Long before the widespread deployment of advanced autonomous agents, Mali's information environment was already contested by state forces and their armed opponents, including Tuareg-led movements and JNIM, as well as by foreign governments and contractors. These actors use interconnected public and private digital networks to advance competing accounts of the conflict. Their narratives draw on locally resonant themes such as sovereignty, betrayal, liberation, and military legitimacy (Africa Center for Strategic Studies, 2024; AFP Africa, 2023; de Bruijn, 2025; Thompson et al., 2022).

If AI swarm-like systems become accessible to actors engaged in Mali's political and armed conflicts, they could exploit precisely the characteristics of its existing communication environment. These networks provide potential channels for diffusion, although whether their structure would help conceal coordination remains to be tested (Schroeder et al., 2023). The forms of citizen journalism described by de Bruijn et al. (2025) could provide stylistic and emotional templates for synthetic accounts, although the ease of reproducing them remains untested. Diaspora networks offer transnational amplification channels. Most importantly, it is already difficult to distinguish between authentic grassroots discourse and orchestrated influence. We expect these dynamics to even become more difficult given that AI makes synthetic participation harder to detect and easier to normalize.

Mali therefore shows how the risks associated with malicious AI swarms could build on observable features of a fragile communication environment. Under conditions of automated manipulation, the same infrastructures that support civic participation and community reporting could accelerate misinformation, intensify factional identities, incite political violence, and weaken democratic reconstruction and peace negotiations (Schroeder et al., 2026). Within this forward-looking analysis, Mali is selected using a most-likely-case rationale: its conflict dynamics, digitally networked communication, and exposure to foreign influence create conditions under which AI-enabled manipulation could have especially serious consequences (Gerring, 2007). The case is not representative of Africa, but it helps identify mechanisms that may also matter in other conflict-affected settings.

%% file: sections/06-case-study-2-ethiopia-as-a-laboratory-of-authoritarian-information-adaptation.tex
\section{Case Study 2: Ethiopia as a Laboratory of Authoritarian Information Adaptation}
\label{sec:case-study-2-ethiopia-as-a-laboratory-of-authoritarian-information-adaptation}

Ethiopia provides a second empirical case that complements Mali by illustrating a different configuration of vulnerability and political agency. Whereas the Malian case highlights how AI-enabled influence operations could exploit a fragmented and conflict-affected information environment characterized by weak gatekeeping and competing external actors, Ethiopia demonstrates how coordinated digital influence can also become embedded within a more organized state strategy of narrative management, regime legitimation, and political mobilization. The Ethiopian case is therefore analytically important not simply because it is multilingual and conflict-affected, but because state and pro-government actors have already demonstrated substantial capacity to organize domestic and diasporic participation in digital influence campaigns.

Research on Ethiopia's recent conflicts shows that social-media infrastructures have become active components of political struggle rather than passive channels through which conflict is reported. During and after the Tigray war, online communication was marked by polarized and often hostile discourse organized around ethnic, national, and geopolitical identities. Orgeret et al. (2026) show how hashtags developed into ``hatetags,'' reinforcing antagonistic group boundaries during the Tigray conflict, while Gessesse and Moges (2026) identify sectarian and hostile rhetoric in English- and Amharic-language Facebook discourse during conflict in Amhara in September--December 2023.

Crucially, however, the Ethiopian case also demonstrates that coordinated influence need not emerge primarily from decentralized or external actors. Kebede and Pohjonen (2025) document how the Ethiopian government and its supporters used Twitter during the 2020--2022 civil war to advance strategic narratives directed simultaneously at domestic and international audiences. Pro-government campaigning portrayed the Tigray People's Liberation Front as an existential threat while framing Western criticism as neo-colonial intervention. These narratives sought to mobilize domestic support for the war and resist external political pressure. The campaign therefore operated not simply as disinformation in the narrow sense, but as a form of participatory propaganda in which supporters---including members of the Ethiopian diaspora---became active participants in reproducing and amplifying state-aligned narratives.

This pattern appears to have developed into a broader strategy of authoritarian adaptation. A recent study of information suppression in Ethiopia describes a shift from online censorship, internet shutdowns, and direct repression only toward complementary strategies aimed at shaping the salience of information in the digital public sphere. These include a government-organized ``media army,'' paid activists and influencers, coordinated messaging, agenda-setting, and the mobilization of diaspora constituencies (Pohjonen et al., 2026a). Rather than merely preventing critical information from circulating, such strategies seek to overwhelm competing narratives, dominate attention, and ensure that political debate remains structured around agendas established by the state (Pohjonen et al., 2026a). Research on China likewise identifies strategic distraction as a means of directing public attention (King et al., 2017).

The Ethiopian case is particularly relevant to AI-enabled influence operations because this informational infrastructure can potentially provide an organizational substrate for increasingly automated coordination. Existing networks of officials, activists, influencers, diaspora supporters, and state-controlled media already perform functions that parallel several elements of the swarm architecture discussed earlier: distributed message production, role differentiation, narrative coordination, audience targeting, and adaptation to changing political events. Generative and agentic AI could reduce the cost of performing these functions, accelerate message variation, segment communication across linguistic and political constituencies, and allow operators to test and adapt narratives more rapidly. In this sense, AI swarms need not emerge as wholly autonomous systems operating outside existing political structures. They may instead be layered onto pre-existing human and institutional networks, producing hybrid influence operations in which automation augments established practices of participatory propaganda and information control.

Recent developments surrounding Ethiopia's campaign for access to the Red Sea illustrate this transition toward AI-assisted political communication. AI-generated images, video, synthetic voice, and music promoting Ethiopia's maritime ambitions have circulated on platforms such as X and TikTok. Some content has been amplified by former government officials and pro-government accounts, using nationalist imagery and narratives of overcoming landlockedness to mobilize support for access to the sea (Tesfaye \& Pohjonen, 2025). The 2026 ARM report similarly identifies AI-generated propaganda concerning the Red Sea as part of the evolving repertoire of state-aligned information management (Pohjonen et al., 2026a). These developments suggest that generative AI may lower the cost of producing emotionally evocative visual propaganda that can support both domestic legitimation and foreign-policy objectives.

The government's appropriation of informational practices associated with verification further complicates this environment. The ARM study documents the co-option of the language of fact-checking within broader state communication strategies, alongside coordinated efforts to control information salience (Pohjonen et al., 2026a, p. 34). This is analytically significant because technologies or institutions associated with information integrity can themselves become incorporated into political struggles over credibility. In such circumstances, future AI-enabled influence operations may not merely generate false content; they could also simulate the language, visual conventions, or authority signals of verification and institutional expertise.

Ethiopia's multilingual environment adds another important dimension. An AI-assisted campaign could segment narratives across Amharic, English, Oromo, Tigrinya, and other linguistic publics, framing the same event through different historical analogies, identity cues, moral vocabularies, or geopolitical narratives. Yet uneven model performance across African languages means that culturally persuasive communication is likely to continue to depend substantially on human linguistic and contextual knowledge. Rather than eliminating human operators, AI may therefore intensify a division of labour in which models provide scale, translation, synthetic media, and rapid variation while politically embedded human actors supply cultural credibility, strategic judgement, and audience knowledge.

Taken together, the Ethiopian case extends the paper's argument in an important direction. The risk posed by malicious AI swarms is not limited to environments in which weak institutions allow synthetic actors to exacerbate instability. In more tightly organized but politically contested settings, AI-enabled coordination may also strengthen the capacity of the regimes to manufacture perceptions of popular support, crowd out dissent, cultivate domestic legitimacy, mobilize diaspora constituencies, and advance foreign-policy objectives. Ethiopia therefore illustrates a second pathway of democratic risk: the integration of increasingly automated influence capabilities into existing structures of authoritarian information control. This differs from Mali's emphasis on exploiting fragmented information environments and shows why the political effects of AI swarms are likely to depend not only on technological capability but on who controls the surrounding organizational and institutional infrastructure. Understood together, the Mali and Ethiopia cases therefore reveal distinct pathways through which AI-enabled influence can interact with conflict, contested legitimacy, and uneven structures of political power.

%% file: sections/07-pathways-of-harms-to-democracy-in-africa.tex
\section{Pathways of Harms to Democracy in Africa}
\label{sec:pathways-of-harms-to-democracy-in-africa}

The two cases suggest that the democratic risks of increasingly automated influence systems arise not from technology alone, but from the interaction between technical capabilities and existing political, institutional, and communicative conditions. Research across multiple African countries shows that political competition can heighten the salience of ethnic identities (Eifert et al., 2010), while state control over digital infrastructure can facilitate election-related internet shutdowns (Freyburg \& Garbe, 2018). These vulnerabilities coexist with substantial civic engagement: Afrobarometer survey data from 39 countries show widespread participation and associations between some forms of engagement, government responsiveness, and democratic performance (Afrobarometer, 2025). Adaptive and scalable disinformation systems could exploit both sides of this environment, intensifying existing divisions while distorting the channels through which citizens make their voices heard.

One of the most significant risks is the fabrication of what appears to be widespread public agreement. When large numbers of synthetic accounts converge on the same narrative, they can create the impression of a broad social consensus where none exists (Schroeder et al., 2026). This illusion of majority support could shift perceptions of what views are common, acceptable, or politically viable (Kunst, Obaidi, et al., 2026). Because political opinion in many African contexts is strongly influenced by cues from social networks and community identity, perceived consensus can itself become persuasive, even if the voices producing it are algorithmically generated.

Such dynamics also threaten public trust in institutions. Many African regimes already contend with skepticism toward electoral commissions, the judiciary, security forces, and the media. Disinformation campaigns that present these institutions as corrupt, incompetent, or foreign-controlled---especially when targeted at specific communities or grievances---could deepen mistrust and weaken the authority of public decision-making. This is a proposed pathway through which disinformation may intensify the democratic risks associated with polarization (McCoy et al., 2018). Once trust erodes, even accurate information may fail to restore legitimacy, producing long-term governance instability.

At the same time, authoritarian regimes in Africa and beyond have shown to use social media in their own campaigns, including blocking social media in response to the removal of government-linked accounts, as during Uganda's 2021 general elections (Lemaire \& Karyeija, 2024), co-opting the language of fact-checking for propaganda purposes (e.g., Alyukov \& Zavadskaya, 2026; Pohjonen et al., 2026a), or using AI-generated materials to push for their own foreign policy goals (Tesfaye \& Pohjonen, 2025). Here, disinformation campaigns can be used to manufacture legitimacy, tighten one's grip on power, and deter political engagement.

In societies where political opinions align with ethnic, regional, or religious identities, these changes in perception can intensify polarization. The conflict narratives described by de Bruijn et al. (2025) suggest a potential target for automated systems capable of mimicking local linguistic and cultural cues: amplifying accounts of betrayal, victimhood, or historical trauma. Repeated amplification of such narratives may normalize hostile rhetoric, harden boundaries between groups, narrow the space for deliberation or compromise, and contribute to escalation under conditions of existing conflict (de Bruijn et al., 2025; McCoy et al., 2018).

Not all manipulation aims to persuade; some is designed to distract or exhaust. By flooding communication channels with distracting claims, sensational stories, or overwhelming volumes of commentary, automated systems can make reliable information harder to identify and divert attention from politically consequential issues (Goldstein et al., 2023; King et al., 2017; Roberts, 2018). Under these conditions, people may withdraw from public discussion because determining what to trust becomes increasingly difficult. Political participation can therefore decline through uncertainty and fatigue, not only through direct silencing. Such disengagement can erode civic life alongside more overt forms of repression.

These risks become particularly acute in electoral and transitional periods. If automated influence systems are able to adjust their messaging in response to public sentiment or polling data, they could mobilize some groups while discouraging others, distort perceptions of electoral fairness, or delegitimize outcomes before they occur (Goldstein et al., 2023). Once the idea takes hold that political processes are inevitably manipulated, the authority of elections and constitutional negotiations can collapse, creating a political vacuum that is difficult to reconstruct.

Taken together, these proposed pathways suggest that the threat of automated disinformation in Africa lies less in dramatic disruption than in the gradual reshaping of everyday political interpretation. Small but continuous shifts in what seems true, who seems credible, and which voices appear to speak for the public can cumulatively alter the foundations of democratic life (McCoy et al., 2018; Schroeder et al., 2026). In fragile contexts, these shifts may unfold quickly and leave little room for recovery once they become entrenched.

%% file: sections/08-governance-and-safeguards.tex
\section{Governance and Safeguards}
\label{sec:governance-and-safeguards}

Addressing the risks posed by increasingly automated disinformation requires a governance strategy that matches the complexity of the systems involved (Helberger, 2020). Because malicious influence campaigns exploit weaknesses in political institutions, platform architectures, and public trust, there is no single intervention that can eliminate the threat. Effective safeguards must instead operate across multiple layers of the information ecosystem---technical, institutional, civic, and transnational (Okolo, 2025).

A first priority is improving the capacity to detect coordinated inauthentic behavior. This need not rely solely on identifying the ``truth'' of claims, which is often politically contested, but rather tracing patterns of organization, timing, narrative alignment, or posting dynamics that indicate artificial coordination (Goldstein et al., 2023). Platforms already possess much of the information required to identify such patterns, yet their moderation strategies have rarely prioritized African contexts. Establishing transparent auditing mechanisms, enabling independent research access to platform data, and committing to regionally relevant detection standards would represent meaningful steps toward reducing manipulation at scale (Okolo, 2025). At the same time, automated moderation is itself politically and technically imperfect: platform-governance research shows that large-scale content moderation depends on contested classifications, uneven enforcement, and trade-offs between safety, expression, transparency, and accountability (Gorwa et al., 2020). These trade-offs also raise questions of informational justice: misinformation-detection systems may distribute benefits and burdens unevenly, constrain participation, misrepresent affected groups, or discount the credibility of particular sources of evidence (Neumann et al., 2022). Importantly, these measures should be implemented in a way that safeguards political speech and avoids enabling state repression---particularly in contexts where governments restrict political expression and access to information (Lutscher, 2025), or of justifying social media shutdowns when their support group is targeted by moderation efforts (Lemaire \& Karyeija, 2024).

Strengthening provenance and identity verification systems could make coordinated manipulation more difficult to conduct at scale. Cryptographic attestation tools, federated identity protocols, and optional ``proof-of-personhood'' systems could help platforms limit the operation of large fleets of synthetic accounts (Adler et al., 2024). However, identity-based verification must be implemented cautiously. In repressive political contexts, anonymity can be essential for activists seeking to reduce the risk of identification and retaliation; evidence from Egypt, for instance, shows that regime opponents used accounts without personal information and deleted mobilization posts to mitigate state repression (Lutscher \& Ketchley, 2023). Safeguards should therefore permit verified but pseudonymous participation, periodic re-verification, and strong legal protections against unmasking protected speakers.

At the societal level, resilience cannot depend solely on technical filtering. Civic actors---journalists, educators, peacebuilding organizations, and community media networks---play a fundamental role in shaping how information is interpreted. Investments in media literacy, local fact-checking networks, community radio partnerships, and translation of verification resources into local languages can help strengthen collective interpretive capacity (Africa Check, n.d.; Okolo, 2025). The aim is not to train citizens to detect every false statement, but to maintain a baseline of epistemic trust and dialogic openness that automated manipulation attempts struggle to collapse.

Governance must also extend beyond the nation-state. Many documented influence operations in Africa involve external state-linked, paramilitary, or commercial actors (Africa Center for Strategic Studies, 2024). Regional bodies such as the African Union, ECOWAS, and IGAD already coordinate election observation, peace mediation, and cybersecurity policy; extending these frameworks to include joint monitoring of AI-enabled influence would provide shared situational awareness. Multilateral ``early warning and verification networks'' could help identify emerging campaigns before they reach political tipping points, particularly during elections, ceasefire negotiations, or constitutional transitions. Such coordination should follow multistakeholder platform-governance principles so that counter-disinformation measures remain transparent, rights-respecting, and accountable rather than becoming tools for political censorship (UNESCO, 2023).

Finally, any governance response must recognize that the line between beneficial and harmful uses of automation is often blurry. AI systems can also be used defensively---to assist journalists, monitor platform manipulation, translate official communication, or expand civic participation. The goal is not to prohibit automated systems, but to shape the conditions under which they operate: to make manipulation more expensive, slower, and easier to be detected, while making authentic civic communication safer, easier, and more durable.

In this sense, safeguarding democracy in Africa, and the Global South more generally,against automated disinformation is not only a matter of regulation or technical defense. It is a broader project of reinforcing the social and institutional infrastructures that sustain public trust (Helberger, 2020). The challenge is to act proactively: before the capabilities of malicious AI swarms mature to the point where intervention becomes significantly more difficult (Schroeder et al., 2026).

%% file: sections/09-towards-resilient-information-ecosystems-for-democracy-and-development.tex
\section{Towards Resilient Information Ecosystems for Democracy and Development}
\label{sec:towards-resilient-information-ecosystems-for-democracy-and-development}

Building resilience against automated disinformation does not mean restoring an imagined past of stable media hierarchies or uncontested public narratives. In many African states, the public sphere has always been plural, negotiated, and shaped by competing visions of authority. The challenge, therefore, is not to suppress complexity but to support the infrastructures that enable democratic deliberation to function within it. Here, we understand resilience as the capacity of societies to absorb disagreement, evaluate competing claims, and maintain trust in the processes that govern collective decision-making.

Strengthening such resilience requires investing in the everyday infrastructures of meaning-making where local journalists, community radio stations, vernacular-language media, fact-checking organizations, and civic educators are central actors (Africa Check, n.d.). Their presence provides anchors of continuity and memory that algorithmic content streams cannot yet replace. Supporting them involves not only funding and training but ensuring the physical safety of media workers, legal protections for civic speech, and economic models that allow independent media to survive without clientelist capture.

Digital development strategies must also move beyond a narrow focus on connectivity as progress. Expanding access without expanding interpretive capacity risks deepening vulnerability to manipulation (Okolo, 2025). Media-literacy initiatives should be developed as situated social practices that reflect local cultural, social, political, and infrastructural conditions (Boshoff \& Fafowora, 2024). Evidence from Kenya and Nigeria also shows that prompting users to consider accuracy can reduce intentions to share COVID-19 misinformation (Offer-Westort et al., 2024). Importantly, such initiatives must avoid framing communities as passive victims of misinformation; their existing relationships, cultural frameworks, and practices of evaluating information should inform programme design (Boshoff \& Fafowora, 2024; Nyabola, 2018).

At the infrastructural level, the long-term goal should be to reduce dependencies on global platforms over which African publics have limited oversight. This concern resonates with critiques of unequal technological dependency in humanitarian settings (Madianou, 2019). Emerging conversations around African digital sovereignty, federated social networking platforms, and community-governed data infrastructures point toward possible alternatives. These approaches do not eliminate the risks of manipulation, but could increase the ability of local institutions to set norms, enforce moderation practices, and design information systems aligned with regional political cultures and development priorities.

Importantly, supporting innovations centered on media independence and the development of alternative models of platform governance may also reduce incentives for African states to resort to network disruptions, cyberattacks, or platform blocks as strategies to disrupt the activities of AI swarms. Research links denial-of-service attacks to elections in more authoritarian regimes (Lutscher et al., 2020) and examines attacks on news websites in Venezuela (Lutscher, 2023a). Other studies examine news-site blocking in Egypt (Lutscher, 2023b) and internet shutdowns in Ethiopia (Chala \& Workneh, 2026). Documented cases include pro-government bot activity in Russia (Stukal et al., 2022) and state-aligned amplification of AI-generated propaganda in Ethiopia (Tesfaye \& Pohjonen, 2025). These established capacities could provide foundations for future swarm-like operations.

Therefore, resilience requires acknowledging that information ecosystems are inseparable from material governance. Trust does not emerge from communication alone. Obadare (2024) argues that rebuilding trust requires governments to respond to citizens through public services and accountable institutions. This link between public trust, citizen engagement, and institutional responsiveness is also consistent with Afrobarometer's finding that electoral participation is higher where citizens are satisfied with democracy, view elections as free, fair, and effective, and feel free to cast their ballots (Afrobarometer, 2025). Efforts to strengthen digital resilience must therefore be linked to broader governance reforms: free, fair and transparent elections, independent oversight institutions, peacebuilding processes, and inclusive development planning. Information integrity is not simply a technical problem but a condition of sociopolitical life.

In this sense, the question posed by emerging AI-driven influence systems is not only how to prevent harm, but how to imagine---and build---public spheres that can withstand it. The task is to ensure that contestation remains compatible with coexistence, that plurality does not collapse into fragmentation, and that the infrastructures of collective sense-making remain strong enough to outlast the technologies that seek to destabilize them. We need to shape these conditions before the capabilities of automated disinformation systems fully mature. What is at stake is not only the integrity of elections or media systems, but the foundations of democratic development itself.

%% file: sections/10-research-agenda-operationalising-ai-swarms-and-testing-their-effects.tex
\section{Research Agenda: Operationalising AI Swarms and Testing Their Effects}
\label{sec:research-agenda-operationalising-ai-swarms-and-testing-their-effects}

A central research priority is to operationalise malicious AI swarms as a graded and empirically testable phenomenon. Studies should distinguish AI-generated content from systems exhibiting multiplicity, persistent roles or identities, coordination around a shared objective, adaptation to feedback, and collective task allocation. These properties should be measured as dimensions rather than binary conditions because AI-enabled influence operations may display different degrees of swarm-like organisation.

Indicators should be examined at the account, network, and system levels. At the account level, researchers can assess persona continuity, interaction histories, role consistency, linguistic and stylistic shifts, and responses to correction (Olejnik, 2026; Olejnik et al., 2026). Network indicators include temporal coordination, task differentiation, repeated interaction, synchronised changes in framing, and movement across platforms. At the system level, evidence may include information sharing, adaptation to interventions, task redistribution, and tactical revision. In our proposed classification, revision of campaign goals would indicate more advanced autonomy but is not required (Haidemariam, 2026).

Operationalisation must also cover hybrid arrangements combining automated agents with political strategists, local-language specialists, influencers, diaspora actors, or account operators. Researchers should examine how labour, information, and decision-making are distributed rather than asking only whether individual accounts are automated. Human editing of AI-generated messages, shared narrative templates, and changes in the division of labour following platform enforcement may help distinguish semi-autonomous swarm behaviour from AI-assisted campaigns that remain under human strategic control.

Simulations such as those introduced by Olejnik et al. (2026) could compare centrally coordinated accounts, adaptive multi-agent systems, and hybrid human--AI campaigns under varying levels of language competence, human involvement, institutional trust, corrective exposure, and platform design. Relevant outcomes include perceived consensus, polarisation, correction reach, institutional trust, political participation, and the transmission of online narratives into offline communication. The aim would be to identify when coordinated synthetic participation becomes influential.

Multilingual validity must be built into this research. Models and detection methods should be evaluated separately by language, dialect, and communication register, with locally knowledgeable speakers assessing idiom, code-switching, historical references, and coded expression. Without such validation, estimates of influence capabilities and defensive performance may be unreliable (Hussen et al., 2026; Ojo et al., 2025).

%% file: sections/11-conclusion.tex
\section{Conclusion}
\label{sec:conclusion}

This paper has argued that the risks posed by emerging AI-driven disinformation campaigns cannot be understood in isolation from the sociopolitical environments in which they operate. The potential development of malicious AI swarms -- coordinated networks of synthetic personas capable of adaptive, sustained influence -- represents a plausible near-future escalation in the automation of digital persuasion. Yet the threat does not lie in technology alone. It emerges from the interaction between AI capabilities and political conditions including contested legitimacy, uneven media infrastructures, polarized digital publics, and differing capacities of state and non-state actors to organize influence.

Mali and Ethiopia illustrate two related but distinct pathways through which AI-enabled influence operations may interact with politically contested information environments. In Mali, a fragmented public sphere shaped by citizen journalists, diaspora actors, political elites, and competing foreign information campaigns creates opportunities for adaptive systems to intensify polarization, fragment shared meaning, and erode confidence in governance. In Ethiopia, by contrast, more organized state-led information strategies show how AI-enabled influence may also be incorporated into participatory propaganda, regime legitimation, diaspora mobilization, and foreign-policy communication. Together, the cases show that the risks of AI swarms are not only confined to weakly governed information environments; they may also arise where political actors possess substantial capacity to organize and direct digital influence. If automated manipulation becomes easier, cheaper, and more contextually tailored, the line between authentic civic discourse and engineered influence may grow increasingly difficult to discern. Language and training-data asymmetries may constrain fully autonomous influence operations while simultaneously weakening moderation, detection, and fact-checking capacity. The most plausible near-term threat may therefore be hybrid, combining AI-generated scale and adaptation with human cultural knowledge, credibility, and strategic judgement.

However, this future is not a self-fulfilling prophecy. The same digital infrastructures that can be weaponized have also expanded access to public life, amplified marginalized voices, and enabled new forms of accountability (Diamond, 2010). The challenge is to reinforce the institutional, civic, and technological conditions that sustain meaningful democratic deliberation. Strengthening provenance systems, increasing transparency in platform moderation, supporting independent and local media, investing in context-specific media literacy, and developing regional mechanisms for monitoring influence operations can all help raise the cost of manipulation while protecting the conditions for genuine political participation. Effective governance must operate across data, model, agent, platform, civic, and regional layers, while protecting anonymity, political expression, and vulnerable users.

Ultimately, resilience will depend not only on defensive measures but on the broader health of democratic institutions, peacebuilding processes, and development frameworks. Automated disinformation is especially destabilizing where trust is already thin, political legitimacy is contested, and state or non-state actors possess the capacity to organize and amplify influence at scale. Efforts to counter emerging influence operations must therefore be linked to wider projects of inclusive governance, social cohesion, and economic justice. Future research should operationalize swarm-like behaviour through observable indicators of persistence, coordination, adaptation, task allocation, and human--AI interaction.

The window of opportunity is present now, while the full capabilities of malicious AI swarms remain emerging rather than entrenched. By anticipating these developments before they become widespread, African governments and their partners have the chance to shape digital ecosystems that are not merely resistant to manipulation but supportive of democratic development, collective reasoning, and shared political futures.

%% file: references.tex
\section*{References}
\addcontentsline{toc}{section}{References}
\begingroup
\small
\begin{referencelist}
\item Adler, S., Hitzig, Z., Jain, S., Brewer, C., Chang, W., DiResta, R., Lazzarin, E., McGregor, S., Seltzer, W., Siddarth, D., Soliman, N., South, T., Spelliscy, C., Sporny, M., Srivastava, V., Bailey, J., Christian, B., Critch, A., Falcon, R., \ldots{} Zick, T. (2024). Personhood credentials: Artificial intelligence and the value of privacy-preserving tools to distinguish who is real online {[}Preprint{]}. arXiv. \url{https://doi.org/10.48550/arXiv.2408.07892}
\item AFP Africa. (2023, November 16). Malian rivals' power struggle spills onto social media. AFP Fact Check. \url{https://factcheck.afp.com/doc.afp.com.343V8PA}
\item Africa Center for Strategic Studies. (2024, March 13). Mapping a surge of disinformation in Africa. \url{https://africacenter.org/spotlight/mapping-a-surge-of-disinformation-in-africa/}
\item Africa Check. (n.d.). Facts in Q\&A format {[}Info Finder{]}. Retrieved October 4, 2026, from \url{https://www.africacheck.org/infofinder/explore-facts}
\item Afrobarometer. (2025). African insights 2025: Citizen engagement, citizen power: Africans claim the promise of democracy. \url{https://www.afrobarometer.org/feature/african-insights-2025/}
\item Alberts, K. (2025). Unravelling the web: Exploring the dynamics of online disinformation and its impact on democracy in Africa. Canadian Journal of African Studies / Revue canadienne des études africaines, 59(2), 333--359. \url{https://doi.org/10.1080/00083968.2025.2491679}
\item Alyukov, M., \& Zavadskaya, M. (2026). ``It's not that simple, we don't know the whole truth'': The effects of pseudo fact checking in wartime Russia. Perspectives on Politics. Advance online publication. \url{https://doi.org/10.1017/S153759272610454X}
\item Badji, S. D., Orgeret, K. S., \& Mutsvairo, B. (2024). An exploratory study of fact-checking practices in conflict and authoritarian contexts. Media and Communication, 12, Article 8698. \url{https://doi.org/10.17645/mac.8698}
\item Beacken, G. D., Trauthig, I. K., \& Woolley, S. C. (2026). Generative AI, propaganda, and digital authoritarianism: Comparative insights from six democratically weakened countries. Big Data \& Society, 13(2), Article 20539517251410064. \url{https://doi.org/10.1177/20539517251410064}
\item Bolsover, G., \& Howard, P. (2019). Chinese computational propaganda: Automation, algorithms and the manipulation of information about Chinese politics on Twitter and Weibo. Information, Communication \& Society, 22(14), 2063--2080. \url{https://doi.org/10.1080/1369118X.2018.1476576}
\item Bosch, T. (2017). Twitter activism and youth in South Africa: The case of \#RhodesMustFall. Information, Communication \& Society, 20(2), 221--232. \url{https://doi.org/10.1080/1369118X.2016.1162829}
\item Bosch, T. E., Mare, A., \& Ncube, M. (2020). Facebook and politics in Africa: Zimbabwe and Kenya. Media, Culture \& Society, 42(3), 349--364. \url{https://doi.org/10.1177/0163443719895194}
\item Boshoff, P., \& Fafowora, B. (2024). Digital media literacy in Africa: Towards a research agenda. African Journalism Studies, 45(4), 259--269. \url{https://doi.org/10.1080/23743670.2025.2478460}
\item Brinkerhoff, J. M. (2009). Digital diasporas: Identity and transnational engagement. Cambridge University Press. \url{https://doi.org/10.1017/CBO9780511805158}

\item Bruls, L., \& de Bruijn, M. (2025). Social media, duress and the Malian conflict. Conflict and Society, 11(1), 75--96. \url{https://doi.org/10.3167/arcs.2025.110106}
\item Chala, E. H., \& Workneh, T. W. (2026). Internet shutdowns in Ethiopia: Discourses of digital sovereignty and information suppression amid political instability. New Media \& Society, 28(9), 4942--4962. \url{https://doi.org/10.1177/14614448251378981}
\item Chesney, R., \& Citron, D. K. (2019). Deep fakes: A looming challenge for privacy, democracy, and national security. California Law Review, 107(6), 1753--1819. \url{https://doi.org/10.15779/Z38RV0D15J}
\item Cinelli, M., De Francisci Morales, G., Galeazzi, A., Quattrociocchi, W., \& Starnini, M. (2021). The echo chamber effect on social media. Proceedings of the National Academy of Sciences, 118(9), Article e2023301118. \url{https://doi.org/10.1073/pnas.2023301118}
\item Conroy-Krutz, J., Amewunou, K., \& Amakoh, K. (2026, April 27). How Africans get their news is changing - how much they get might not be (Afrobarometer Dispatch No. 1173). Afrobarometer. \url{https://www.afrobarometer.org/publication/ad1173-how-africans-get-their-news-is-changing-how-much-they-get-might-not-be/}
\item de Bruijn, M. (2025, October 14). Digital communication as a weapon: The case of Mali. Carnegie Endowment for International Peace. \url{https://carnegieendowment.org/research/2025/10/digital-communication-as-a-weapon-mali}
\item de Bruijn, M., Mutsvairo, B., Bruls, L., Cissé, M. G., Langguth, J., Orgeret, K. S., Badji, S. D., Moges, M. A., \& Schroeder, D. T. (2025). ``Aren't we all journalists?'' Citizen journalism, disinformation and the weaponization of social media in conflict torn Mali. Journalism, 26(5), 1027--1046. \url{https://doi.org/10.1177/14648849241312743}
\item Dembele, K., Ballo, I., Kone, B., Sissoko, E. F., \& Dembele, A. (2024). Réseaux sociaux au Mali : Entre mobilisation politique et mécanismes de censure. Revue Internationale des Sciences de Gestion, 7(1), 267--292. \url{https://doi.org/10.5281/zenodo.10582646}
\item Diamond, L. (2010). Liberation technology. Journal of Democracy, 21(3), 69--83. \url{https://doi.org/10.1353/jod.0.0190}
\item Dylan, H., \& Grossfeld, E. (2025). Revisionist future: Russia's assault on large language models, the distortion of collective memory, and the politics of eternity. Dialogues on Digital Society, 1(3), 407--412. \url{https://doi.org/10.1177/29768640251377941}
\item Eady, G., Paskhalis, T., Zilinsky, J., Bonneau, R., Nagler, J., \& Tucker, J. A. (2023). Exposure to the Russian Internet Research Agency foreign influence campaign on Twitter in the 2016 US election and its relationship to attitudes and voting behavior. Nature Communications, 14, Article 62. \url{https://doi.org/10.1038/s41467-022-35576-9}
\item Eifert, B., Miguel, E., \& Posner, D. N. (2010). Political competition and ethnic identification in Africa. American Journal of Political Science, 54(2), 494--510. \url{https://doi.org/10.1111/j.1540-5907.2010.00443.x}
\item European External Action Service. (2023). 1st EEAS report on foreign information manipulation and interference threats: Towards a framework for networked defence. \url{https://www.eeas.europa.eu/sites/default/files/documents/2023/EEAS-DataTeam-ThreatReport-2023..pdf}
\item Fisher, J., Gadjanova, E., \& Hitchen, J. (2024). WhatsApp and political communication in West Africa: Accounting for differences in parties' organization and message discipline online. Party Politics, 30(5), 934--948. \url{https://doi.org/10.1177/13540688231188690}
\item Freedom House. (2024). Freedom in the World 2024: Mali. \url{https://freedomhouse.org/country/mali/freedom-world/2024}
\item Freyburg, T., \& Garbe, L. (2018). Blocking the bottleneck: Internet shutdowns and ownership at election times in Sub-Saharan Africa. International Journal of Communication, 12, 3896--3916. \url{https://ijoc.org/index.php/ijoc/article/view/8546}
\item Gerring, J. (2007). Is there a (viable) crucial-case method? Comparative Political Studies, 40(3), 231--253. \url{https://doi.org/10.1177/0010414006290784}
\item Gessesse, K. B., \& Moges, M. A. (2026). Social media discourses amidst ethnopolitical extremism and conflict: The case of Ethiopia. Discourse Studies, 28(3), 408--433. \url{https://doi.org/10.1177/14614456251374251}
\item Goldstein, J. A., Sastry, G., Musser, M., DiResta, R., Gentzel, M., \& Sedova, K. (2023). Generative language models and automated influence operations: Emerging threats and potential mitigations {[}Preprint{]}. arXiv. \url{https://doi.org/10.48550/arXiv.2301.04246}
\item Gondwe, G., Madrid-Morales, D., Tully, M., \& Wasserman, H. (2026). Misinformation and digital inequalities: Comparing how different demographic groups get exposed to and engage with false information. Mass Communication and Society, 29(1), 1--15. \url{https://doi.org/10.1080/15205436.2025.2474139}
\item Gorwa, R., Binns, R., \& Katzenbach, C. (2020). Algorithmic content moderation: Technical and political challenges in the automation of platform governance. Big Data \& Society, 7(1), 1--15. \url{https://doi.org/10.1177/2053951719897945}
\item Graphika, \& Stanford Internet Observatory. (2020, December 15). More-Troll Kombat: French and Russian influence operations go head to head targeting audiences in Africa. \url{https://public-assets.graphika.com/reports/graphika_stanford_report_more_troll_kombat.pdf}
\item Grub, M. F., \& Humprecht, E. (2025). Defining the role(s) of AI in disinformation research---A systematic review. International Journal of Communication, 19, 3577--3601. \url{https://ijoc.org/index.php/ijoc/article/view/24780}
\item GSMA. (2024). The mobile economy Sub-Saharan Africa 2024. \url{https://event-assets.gsma.com/pdf/GSMA_ME_SSA_2024_Web.pdf}
\item Haidemariam, T. (2026). From the logic of coordination to goal-directed reasoning: The agentic turn in artificial intelligence. Frontiers in Artificial Intelligence, 8, Article 1728738. \url{https://doi.org/10.3389/frai.2025.1728738}
\item Haidemariam, T., \& Gran, A.-B. (2026). On the problems of training generative AI: Towards a hybrid approach combining technical and non-technical alignment strategies. AI \& Society, 41(1), 629--654. \url{https://doi.org/10.1007/s00146-025-02445-0}
\item Haidemariam, T., \& Zheng, X. (2026). Synthetic embodiment: A framework for AI-generated identity cues in virtual reality and immersive environments. Frontiers in Virtual Reality, 7, Article 1860842. \url{https://doi.org/10.3389/frvir.2026.1860842}
\item Helberger, N. (2020). The political power of platforms: How current attempts to regulate misinformation amplify opinion power. Digital Journalism, 8(6), 842--854. \url{https://doi.org/10.1080/21670811.2020.1773888}
\item Human Rights Watch. (2024). World report 2024: Mali. \url{https://www.hrw.org/world-report/2024/country-chapters/mali}
\item Huntley, S. (2022, July 29). TAG bulletin: Q2 2022. Google. \url{https://blog.google/threat-analysis-group/tag-bulletin-q2-2022/}
\item Hussen, K. Y., Sewunetie, W. T., Ayele, A. A., Imam, S. H., Alemu, E. N., Muhammad, S. H., \& Yimam, S. M. (2026). The state of large language models for African languages: Progress and challenges. In H. Haddad, A. N. Kahira, S. Bourhim, I. E. Olatunji, L. Makhafola, \& C. Mwase (Eds.), DLI 2025 Research Track (Proceedings of Machine Learning Research, Vol. 302, pp. 1--27). PMLR. \url{https://proceedings.mlr.press/v302/hussen26a.html}
\item International Telecommunication Union. (2023). Measuring digital development: Facts and figures 2023. \url{https://www.itu.int/itu-d/reports/statistics/facts-figures-2023/}
\item International Telecommunication Union. (2025). Measuring digital development: Facts and figures 2025. \url{https://www.itu.int/itu-d/reports/statistics/facts-figures-2025/}
\item Joshi, P., Santy, S., Budhiraja, A., Bali, K., \& Choudhury, M. (2020). The state and fate of linguistic diversity and inclusion in the NLP world. In D. Jurafsky, J. Chai, N. Schluter, \& J. Tetreault (Eds.), Proceedings of the 58th Annual Meeting of the Association for Computational Linguistics (pp. 6282--6293). Association for Computational Linguistics. \url{https://doi.org/10.18653/v1/2020.acl-main.560}
\item Katz, E., \& Lazarsfeld, P. F. (1955). Personal influence: The part played by people in the flow of mass communications. Free Press.
\item Kebede, A. T., \& Pohjonen, M. (2025). `Adwa, not AGOA': Digital propaganda and strategic narratives during the Ethiopian civil war of 2020 to 2022. Media, War \& Conflict. Advance online publication. \url{https://doi.org/10.1177/17506352251403008}
\item King, G., Pan, J., \& Roberts, M. E. (2017). How the Chinese government fabricates social media posts for strategic distraction, not engaged argument. American Political Science Review, 111(3), 484--501. \url{https://doi.org/10.1017/S0003055417000144}
\item Kokutse, F., \& Thompson, J. (2024, May 19). Companies are trying to attract more smartphone users across Africa. But there are risks. Associated Press. \url{https://apnews.com/article/e6cbfeb2b974c688a0a298f5e624538f}
\item Kunst, J. R., Bierwiaczonek, K., Cha, M., Ebrahimi, O. V., Fawcett-Atkinson, M., Følstad, A., Gollwitzer, A., Köbis, N., Marcus, G., Roozenbeek, J., Schroeder, D. T., Van Bavel, J. J., van der Linden, S., White, R., \& Wilhelmsen, L. L. (2026). Puppets or partners? Governing cyborg propaganda in the digital public square {[}Preprint{]}. arXiv. \url{https://doi.org/10.48550/arXiv.2602.13088}
\item Kunst, J. R., Obaidi, M., Gollwitzer, A., Brandtzæg, P. B., Hinrichs, Y., Saini, N., \& Schroeder, D. T. (2026). Intelligent systems, vulnerable minds: A framework for radicalization to violence in the age of AI. Personality and Social Psychology Review, 30(3), 395--426. \url{https://doi.org/10.1177/10888683261430089}
\item Lemaire, P., \& Karyeija, G. K. (2024). Social media and parliamentary candidates in Uganda. The Journal of Modern African Studies, 62(4), 337--358. \url{https://doi.org/10.1017/S0022278X25000102}

\item Lutscher, P. M. (2023a). Hot topics: Denial-of-service attacks on news websites in autocracies. Political Science Research and Methods, 11(4), 696--711. \url{https://doi.org/10.1017/psrm.2021.68}

\item Lutscher, P. M. (2023b). When censorship works: Exploring the resilience of news websites to online censorship. British Journal of Political Science, 53(4), 1342--1350. \url{https://doi.org/10.1017/S0007123422000722}
\item Lutscher, P. M. (2025). Censorship. In A. Nai, M. Grömping, \& D. Wirz (Eds.), Elgar encyclopedia of political communication (Vol. 1, pp. 180--183). Edward Elgar Publishing. \url{https://doi.org/10.4337/9781035301447.vol1.00050}
\item Lutscher, P. M., \& Ketchley, N. (2023). Online repression and tactical evasion: Evidence from the 2020 Day of Anger protests in Egypt. Democratization, 30(2), 325--345. \url{https://doi.org/10.1080/13510347.2022.2140798}
\item Lutscher, P. M., Weidmann, N. B., Roberts, M. E., Jonker, M., King, A., \& Dainotti, A. (2020). At home and abroad: The use of denial-of-service attacks during elections in nondemocratic regimes. Journal of Conflict Resolution, 64(2--3), 373--401. \url{https://doi.org/10.1177/0022002719861676}
\item Madianou, M. (2019). Technocolonialism: Digital innovation and data practices in the humanitarian response to refugee crises. Social Media + Society, 5(3), Article 2056305119863146. \url{https://doi.org/10.1177/2056305119863146}
\item Mare, A. (2020). State-ordered internet shutdowns and digital authoritarianism in Zimbabwe. International Journal of Communication, 14, 4244--4263. \url{https://ijoc.org/index.php/ijoc/article/view/11494}
\item McCoy, J., Rahman, T., \& Somer, M. (2018). Polarization and the global crisis of democracy: Common patterns, dynamics, and pernicious consequences for democratic polities. American Behavioral Scientist, 62(1), 16--42. \url{https://doi.org/10.1177/0002764218759576}
\item Meta. (2021a, January 12). December 2020 coordinated inauthentic behavior report. \url{https://about.fb.com/news/2021/01/december-2020-coordinated-inauthentic-behavior-report/}
\item Meta. (2021b, February 9). January 2021 coordinated inauthentic behavior report. \url{https://about.fb.com/news/2021/02/january-2021-coordinated-inauthentic-behavior-report/}
\item Mutsvairo, B., Orgeret, K. S., de Bruijn, M., Bruls, L., Schroeder, D. T., Galy Cissé, M., Langguth, J., Alemayehu Moges, M., \& Dialimpa Badji, S. (2026). The messenger (not the medium) is the message? Uncovering journalists' perceptions of digital disinformation in conflict countries. Journalism Studies, 27(2), 228--248. \url{https://doi.org/10.1080/1461670X.2025.2575499}
\item Mutsvairo, B., \& Salgado, S. (2022). Is citizen journalism dead? An examination of recent developments in the field. Journalism, 23(2), 354--371. \url{https://doi.org/10.1177/1464884920968440}
\item Nanfuka, J., Kapiyo, V., Mabutho, V., \& Wakabi, W. (2024). A tapestry of actors, attitudes, and impact: Countering disinformation in Africa. Bertelsmann Stiftung. \url{https://doi.org/10.11586/2024059}
\item Neumann, T., De-Arteaga, M., \& Fazelpour, S. (2022). Justice in misinformation detection systems: An analysis of algorithms, stakeholders, and potential harms. In Proceedings of the 2022 ACM Conference on Fairness, Accountability, and Transparency (pp. 1504--1515). Association for Computing Machinery. \url{https://doi.org/10.1145/3531146.3533205}
\item Nyabola, N. (2018). Digital democracy, analogue politics: How the Internet era is transforming politics in Kenya. Zed Books. \url{https://doi.org/10.5040/9781350219656}
\item Obadare, E. (2024). How African governments can regain the trust of their citizens. In A. U. Ordu (Ed.), Foresight Africa: Top priorities for the continent in 2024 (pp. 178--180). Brookings Institution. \url{https://www.brookings.edu/wp-content/uploads/2024/01/ForesightAfrica2024.pdf}
\item Offer-Westort, M., Rosenzweig, L. R., \& Athey, S. (2024). Battling the coronavirus `infodemic' among social media users in Kenya and Nigeria. Nature Human Behaviour, 8(5), 823--834. \url{https://doi.org/10.1038/s41562-023-01810-7}
\item Ojo, J., Ogundepo, O., Oladipo, A., Ogueji, K., Lin, J., Stenetorp, P., \& Adelani, D. I. (2025). AfroBench: How good are large language models on African languages? In W. Che, J. Nabende, E. Shutova, \& M. T. Pilehvar (Eds.), Findings of the Association for Computational Linguistics: ACL 2025 (pp. 19048--19095). Association for Computational Linguistics. \url{https://doi.org/10.18653/v1/2025.findings-acl.976}
\item Okolo, C. T. (2025). African democracy in the era of generative disinformation: Challenges and countermeasures against AI-generated propaganda. In Africa Just AI conference proceedings (pp. 10--19). Research ICT Africa. \url{https://researchictafrica.net/wp-content/uploads/2025/06/JUST_AI_Conference_proceedings_RIA_2025.pdf}
\item Olejnik, L. (2026). Artificial intelligence propaganda factories with language models. Applied Cybersecurity \& Internet Governance, 5(1). \url{https://doi.org/10.60097/ACIG/215416}
\item Olejnik, L., Dong, W., Kunst, J. R., Riemer-Sørensen, S., Herb, T., Cha, M., \& Schroeder, D. T. (2026). IO Factory: Simulating AI-enabled influence campaigns at scale {[}Preprint{]}. arXiv. \url{https://doi.org/10.48550/arXiv.2608.10920}
\item Orgeret, K. S., Mutsvairo, B., de Bruijn, M., Schroeder, D. T., Badji, S. D., \& Moges, M. A. (2026). Hashtags, hatetags and social media campaigns in Ethiopia's Tigray conflict. Information, Communication \& Society, 29(3), 850--868. \url{https://doi.org/10.1080/1369118X.2025.2533314}
\item Orlando, G. M., Ye, J., La Gatta, V., Saeedi, M., Moscato, V., Ferrara, E., \& Luceri, L. (2026). Emergent coordinated behaviors in networked LLM agents: Modeling the strategic dynamics of information operations. In Proceedings of the ACM Web Conference 2026 (pp. 4805--4816). Association for Computing Machinery. \url{https://doi.org/10.1145/3774904.3792580}
\item Pohjonen, M., Kebede, A., \& Aalen, L. (2026a, March). Information suppression in Ethiopia {[}Working paper, Deliverable D5.1{]}. ARM Project. \url{https://www.arm-project.eu/publication/information-suppression-in-ethiopia/}
\item Pohjonen, M., Kebede, A., \& Aalen, L. (2026b, March). Shaping information salience in Ethiopia (ARM Policy Brief 7/26). ARM Project. \url{https://www.arm-project.eu/publication/shaping-information-salience-in-ethiopia/}
\item Roberts, M. E. (2018). Censored: Distraction and diversion inside China's Great Firewall. Princeton University Press. \url{https://doi.org/10.23943/9781400890057}
\item Rydzak, J., Karanja, M., \& Opiyo, N. (2020). Dissent does not die in darkness: Network shutdowns and collective action in African countries. International Journal of Communication, 14, 4264--4287. \url{https://ijoc.org/index.php/ijoc/article/view/12770}
\item Schroeder, D. T., Cha, M., Baronchelli, A., Bostrom, N., Christakis, N. A., Garcia, D., Goldenberg, A., Kyrychenko, Y., Leyton-Brown, K., Lutz, N., Marcus, G., Menczer, F., Pennycook, G., Rand, D. G., Ressa, M., Schweitzer, F., Song, D., Summerfield, C., Tang, A., \ldots{} Kunst, J. R. (2026). How malicious AI swarms can threaten democracy. Science, 391(6783), 354--357. \url{https://doi.org/10.1126/science.adz1697}
\item Schroeder, D. T., de Bruijn, M., Bruls, L., Moges, M. A., Badji, S. D., Fritz, N., Cisse, M. G., Langguth, J., Mutsvairo, B., \& Orgeret, K. S. (2023). Social media in the Global South: A network dataset of the Malian Twittersphere. Journal of Data Mining \& Digital Humanities, 2023, Article 11246. \url{https://doi.org/10.46298/jdmdh.11246}
\item Shanahan, M., \& Bahia, K. (2024). The state of mobile internet connectivity 2024. GSMA. \url{https://prod-cms.gsmaintelligence.com/research-file-download?confluenceId=88244644&filename=231024-Mobile-Internet-Connectivity-2024.pdf}
\item Stukal, D., Sanovich, S., Bonneau, R., \& Tucker, J. A. (2022). Why botter: How pro-government bots fight opposition in Russia. American Political Science Review, 116(3), 843--857. \url{https://doi.org/10.1017/S0003055421001507}
\item Tesfaye, A., \& Pohjonen, M. (2025, October 9). AI and visual propaganda are being used to stoke tensions as Ethiopia eyes Eritrea's Red Sea port. Global Voices. \url{https://globalvoices.org/2025/10/09/escaping-the-geographic-prison-ai-and-visual-propaganda-in-ethiopias-red-sea-discourse/}
\item Thompson, J., Doxsee, C., \& Bermudez, J. S., Jr. (2022, February 2). Tracking the arrival of Russia's Wagner Group in Mali. Center for Strategic and International Studies. \url{https://www.csis.org/analysis/tracking-arrival-russias-wagner-group-mali}
\item UNESCO. (2023). Guidelines for the governance of digital platforms: Safeguarding freedom of expression and access to information through a multistakeholder approach. \url{https://unesdoc.unesco.org/ark:/48223/pf0000387339}
\item United Nations. (2015). Transforming our world: The 2030 Agenda for Sustainable Development (A/RES/70/1). \url{https://sdgs.un.org/2030agenda}
\item Vaswani, A., Shazeer, N., Parmar, N., Uszkoreit, J., Jones, L., Gomez, A. N., Kaiser, Ł., \& Polosukhin, I. (2017). Attention is all you need. Advances in Neural Information Processing Systems, 30, 5998--6008. \url{https://proceedings.neurips.cc/paper/2017/hash/3f5ee243547dee91fbd053c1c4a845aa-Abstract.html}
\item Wack, M., Ehrett, C., Linvill, D., \& Warren, P. (2025). Generative propaganda: Evidence of AI's impact from a state-backed disinformation campaign. PNAS Nexus, 4(4), pgaf083. \url{https://doi.org/10.1093/pnasnexus/pgaf083}
\item Wack, M., Linvill, D., \& Warren, P. (2024). Old despots, new tricks: An AI-empowered pro-Kagame/RPF coordinated influence network on X (Media Forensics Hub Reports No. 5). Clemson University. \url{https://open.clemson.edu/mfh_reports/5/}
\item Waight, H., Yang, E., Yuan, Y., Messing, S., Roberts, M. E., Stewart, B. M., \& Tucker, J. A. (2026). State media control influences large language models. Nature, 655(8123), 685--693. \url{https://doi.org/10.1038/s41586-026-10506-7}
\item Wang, L., Ma, C., Feng, X., Zhang, Z., Yang, H., Zhang, J., Chen, Z., Tang, J., Chen, X., Lin, Y., Zhao, W. X., Wei, Z., \& Wen, J. (2024). A survey on large language model based autonomous agents. Frontiers of Computer Science, 18(6), Article 186345. \url{https://doi.org/10.1007/s11704-024-40231-1}
\item Wardle, C., \& Derakhshan, H. (2017). Information disorder: Toward an interdisciplinary framework for research and policy making (DGI(2017)09). Council of Europe. \url{https://rm.coe.int/information-disorder-report/1680764666}
\item Wasserman, H., \& Madrid-Morales, D. (Eds.). (2022). Disinformation in the Global South. Wiley-Blackwell. \url{https://doi.org/10.1002/9781119714491}
\item Wilson, J. (2026, August 26). Fake US thinktank set up and funded by Israel sought to game AI for propaganda. The Guardian. \url{https://www.theguardian.com/world/2026/aug/26/fake-thinktank-israel-ai-propaganda}
\item Young, L. E. (2023). Mobilization under threat: Emotional appeals and pro-opposition political participation online. Political Behavior, 45(2), 445--468. \url{https://doi.org/10.1007/s11109-021-09711-z}

\end{referencelist}
\endgroup